\documentclass[twocolumn,showpacs,amssymb,nobibnotes, aps,pra]{revtex4}

\usepackage{enumerate,amsmath}
\usepackage{graphicx}
\begin{document}

\title{Entanglement Purification with Double Selection}%

\author{Keisuke Fujii}
\author{Katsuji Yamamoto}
\affiliation{
Department of Nuclear Engineering, Kyoto University, Kyoto 606-8501, Japan}

\date{\today}
\begin{abstract}
We investigate efficient entanglement purification
through double selection process.
This method works with higher noise thresholds
for the communication channels and local operations,
and achieves higher fidelity of purified states.
Furthermore it
provides a yield
comparable to the usual protocol with single selection. 
It is shown by general considerations that 
the double selection is optimal
to remove the first-order errors,
achieving the upper bound on the fidelity of purified states
in the low noise regime. 
The double selection is also applied to
purification of multi-partite entanglement
such as two-colorable graph states.
\end{abstract}

\pacs{03.67.Hk, 03.67.Pp, 03.67.-a}

\maketitle

\section{Introduction}
\label{Introduction}
Recently a number of protocols based on entanglement
have been developed in quantum communication and computation.
For example, quantum teleportation, superdense coding, quantum cryptography
and quantum repeater
employ bipartite entanglement 
\cite{Bennett93,Bennett92,Ekert91,Briegel98}.
Cluster state computation, quantum error correction
and multiparty cryptography
utilize multi-partite entanglement 
\cite{Raussen01,Shor96,Chen04}.
The performance of these entanglement-based protocols
highly depends on the fidelity of entangled states.
That is, high fidelity entangled states
are essential for secure communication and reliable computation.
In this viewpoint, it is a very important task
to prepare and share high fidelity entangled states.

A way to share high fidelity entangled states
via noisy communication channels is known
as entanglement purification 
\cite{BBPSSW96,DEJMPS96}.
This protocol is proposed originally to share EPR states,
and then extended for a large class of multi-partite entangled states,
including the GHZ states, 
two-colorable graph states,
stabilizer states
and W states 
\cite{Murao98,ADB0305,Glancy06,Kruszynska06a,Kruszynska06b,Miyake05}.
In a bipartite situation with noisy channels and perfect local operations,
we pre-purify initial states with a recurrence protocol,
which gives a high threshold for the noise of the communication channel
but a low yield of purified states.
Then, a hashing protocol may be implemented
to get pure entangled states with a nonzero yield.
The hashing protocol, however, breaks down
as soon as local operations become slightly imperfect.
The entanglement purification in such a situation
is first analyzed in the context of quantum repeater \cite{Briegel98},
where the usual recurrence protocol \cite{BBPSSW96,DEJMPS96}
is applied with imperfect local operations.
In this situation, the fidelity of purified states is limited,
and noise thresholds are required
for the local operations as well as the communication channels.
This is the most distinct point 
from the case with perfect local operations.
Thus, in order to realize entanglement-based protocols
using practical devices, which inevitably have imperfections,
we need to develop efficient purification methods,
which work well even with noisy communication channels 
and imperfect local operations.

In this paper 
we investigate an efficient entanglement purification protocol
with more accurate post-selection through double verification process.
This method with double selection enables us to achieve
considerably high fidelity of purified states
even with actual imperfection. 
Furthermore
it works with higher noise thresholds
for the communication channels and local operations
compared with the single-selection scheme \cite{BBPSSW96,DEJMPS96}.
It may be considered that 
the elaborate post-selection
consumes more resources
to decrease the yield of purification.
However, this is not necessarily the case.
The present method with double selection provides
a yield comparable to the single-selection scheme, 
especially when higher fidelity states are required
with noisier local operations.
It is really shown that the double selection is optimal
to remove the first-order errors, achieving the upper bound 
on the fidelity of purified states in the low noise regime.
The double selection 
is also applied to purification 
of multi-partite entanglement
such as two-colorable graph states.
As an example,
we numerically investigate the performance 
for a CSS code (Steane's 7-qubit code),
and compare it with 
a multi-partite purification protocol with single selection \cite{ADB0305}.
As the double selection
can achieve considerably high fidelity
under noisy channels and operations,
it is indeed profitable for quantum computation
as well as quantum communication.

The rest of the paper is organized as follow.
In Sec. \ref{Bipartite} we investigate 
the double selection
in the bipartite entanglement purification.
The performance of the double-selection protocol
is analyzed in detail 
and compared with the usual protocol 
with single selection.
The optimality of the double selection is also discussed
in the low noise regime.
The double selection is then applied 
to the multi-partite entanglement purification
in Sec. \ref{Multi-partite},
where as an example
the purification of the Steane's 7-qubit code is numerically investigated.
Sec. \ref{conclusion} is devoted to conclusion.

\section{Bipartite entanglement purification}
\label{Bipartite}
\subsection{Single selection} 
\begin{figure}
\centering
\scalebox{.45}{\includegraphics*[1cm,1cm][20cm,8cm]{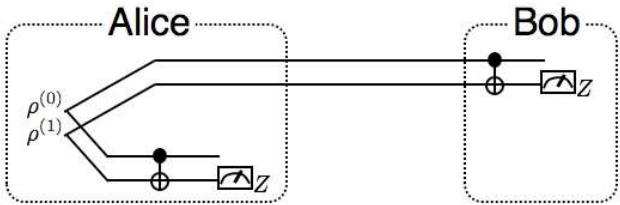}}
\caption{Bipartite entanglement purification with single selection.}
\label{single}
\end{figure}

We first review the usual recurrence protocol
for purification with single selection 
\cite{BBPSSW96,DEJMPS96}.
Suppose 
Alice and Bob share
EPR pairs,
which are decohered via a noisy quantum channel,
and purify them 
by using noisy C-Not gates and a classical channel.
The recurrence protocol 
\cite{BBPSSW96,DEJMPS96} 
consists of
two copies of an EPR pair, 
a bilateral C-Not gate and a bilateral measurement
at each round of purification (see Fig. \ref{single}).
Here, a bilateral operation
means a tensor product of two identical operations
which are simultaneously implemented by Alice and Bob.
The purification procedure is specifically described as follows:
\begin{enumerate}[(i)]
\item
Alice and Bob share two identical EPR pairs
$ \rho^{(0)} $ and $ \rho^{(1)} $
through a noisy quantum channel.

\item
They operate a bilateral C-Not gate
for $ \rho^{(0)} $ and $ \rho ^{(1)} $
as the control and target qubits, respectively.

\item
They bilaterally measure $ \rho^{(1)} $
in the $ Z $ basis $\{ |0\rangle , |1\rangle \}$,
and obtain the measurement outcomes 
$m_{a}$ (Alice) and $m_{b}$ (Bob).

\item
They communicate these measurement outcomes to each other.
Then, they keep $\rho^{(0)}$ 
if the measurement outcomes coincide as $m_{a}=m_{b}$. 
Otherwise, they discard $\rho^{(0)}$.  
\end{enumerate} 
Alice and Bob iterate the procedures (ii)-(iv)
by using the output states which survive the selection in (iv)
as the input states for the next round of purification
where the $ X $ and $ Z $ bases of their reference frames
are exchanged with a Hadamard transformation \cite{DEJMPS96}.
As seen in the above, 
a single bilateral operation
determines whether $\rho^{(0)}$ should be kept or discarded at each round.
That is, the purification procedure is made with single selection.

The noisy EPR pairs $\rho ^{(0)}$ and $\rho ^{(1)}$
are given as two copies of a Bell diagonal state $\rho$,
\begin{equation}
\rho ^{(0)} = \rho ^{(1)} =\rho
= \sum _{j=0}^{3} F_{j} \phi _{j}  ,
\end{equation}
where
\begin{eqnarray}
\phi_{j} &=& | \phi_{j} \rangle \langle \phi_{j} | ,
\\
| \phi_{j} \rangle &=& 
\sigma_{j} \otimes \sigma_{0} (|00 \rangle + |11\rangle) /\sqrt{2}
\end{eqnarray}
($ \sigma_{0} = I $).
In the case with ideal (perfect) local operations, 
the above purification procedure with single selection
provides the output states of $\rho ^{(0)}$
depending on the initial states of $\rho^{(1)}$ as
\begin{equation}
\begin{array}{c|cccc}
 \rho^{(0)} \diagdown \rho^{(1)} 
 & \phi _{0}^{(1)} & \phi ^{(1)}_{1} & \phi^{(1)} _{2} & \phi^{(1)} _{3}
 \\
 \hline
 \phi _{0}^{(0)} & \phi ^{(0)}_{0} & \times &  \times & \phi ^{(0)}_{3}
 \\
 \phi _{1}^{(0)} & \times & \phi ^{(0)}_{1} & \phi ^{(0)}_{2} & \times
 \\
 \phi _{2}^{(0)} &  \times & \phi ^{(0)}_{2} & \phi ^{(0)}_{2} & \times
 \\
 \phi _{3}^{(0)} &  \phi ^{(0)}_{3} & \times & \times & \phi ^{(0)}_{0}
\end{array}
\nonumber
\end{equation}
where $\times$ denotes the output states discarded 
by the post-selection.
Alice and Bob are to distill $|\phi _{0}\rangle$
with the largest probability $F_{0}$
(with a suitable basis transformation if necessary).
For the most likely input state
$\phi^{(0)} _{0} \otimes \phi^{(1)} _{0}$
with the probability $F_{0}^2$, 
the purification procedure 
gives the desired output state $\phi ^{(0)} _{0}$.
On the other hand, the input state $\phi ^{(0)}_{0} \otimes \phi^{(1)} _{3}$
with the probability $F_{0}F_{3}$
results in a wrong output state $\phi ^{(0)}_{3}$,
and so on. 
This may be viewed as the $\sigma _{3}$ error
of $\phi ^{(1)}_{3} = ( \sigma _{3} \otimes \sigma _{0})
\phi ^{(1)}_{0} ( \sigma _{3} \otimes \sigma _{0})$ in $\rho ^{(1)}$
is propagated to $\rho ^{(0)}$ through the purification procedure.
In this way,
the purification procedure provides a map 
$\rho  \rightarrow \rho '$ 
for the Bell diagonal states.
This is even the case
when the Pauli noise is introduced for the local operations,
which is described later.
This purification map can be represented by
a $R^{4} \rightarrow R^{4}$ map $\mathcal{S}$ 
from the input state-vector 
$\mathbf{F}=(F_{0},F_{1},F_{2},F_{3})$
to the output state-vector $\mathbf{F}' =(F'_{0},F'_{1},F'_{2},F'_{3})$ as
\begin{equation}
\mathbf{F}' = \mathcal{S}(\mathbf{F}),
\end{equation} 
or in terms of a transition probability tensor $S_{i}^{jk}$ as
\begin{equation}
F'_{i} =\frac{1}{p_S(\mathbf{F})} \sum _{jk} S_{i}^{jk} F_{j}F_{k} ,
\end{equation}
where
\begin{equation}
p_{\mathcal{S}}(\mathbf{F})=  \sum _{ijk} S_{i}^{jk} F_{j}F_{k} 
\label{success_prob_single}
\end{equation}
is  the success probability responsible
for the normalization $\displaystyle\sum _{i}F'_{i}=1$.
The transition probability tensor 
$S^{jk}_{i}$ is given
including the noise parameters for the local operations
and the exchange of the reference frames at each round
(see Appendix A for the detail).
Then the maximum achievable 
fidelity of purified states
is determined as an appropriate
fixed point of the map 
$\mathcal{S}$.

\subsection{Double selection}
\begin{figure}
\centering
\scalebox{.45}{\includegraphics*[0cm,0cm][20cm,10cm]{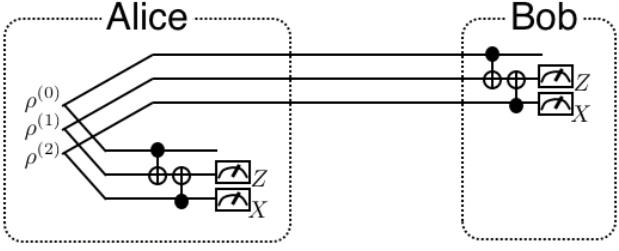}}
\caption{Bipartite entanglement purification with double selection.}
\label{double}
\end{figure}

We now present
the entanglement purification with double selection.
The double-selection protocol
consists of
three copies of EPR pairs, 
two bilateral C-Not gates 
and two bilateral measurements
at each round of purification (see Fig. \ref{double}),
as described in the following:
\begin{enumerate}[(i)]
\item
Alice and Bob share three identical EPR pairs
$ \rho^{(0)} $, $ \rho^{(1)} $ and $\rho^{(2)}$
through a noisy quantum channel.

\item
They operate a bilateral C-Not gate
for $ \rho^{(0)} $ and $ \rho ^{(1)} $
as the control and target qubits, respectively.

\item
Next they operate a bilateral C-Not gate
for $ \rho^{(2)} $ and $ \rho ^{(1)} $
as the control and target qubits, respectively.

\item
They bilaterally measure $ \rho^{(1)} $ and $\rho^{(2)}$
in the $ Z $ and $ X $ bases respectively,
and obtain the measurement outcomes 
$m^{(1)}_{a}$, $m^{(2)}_{a}$ (Alice) 
and $m^{(1)}_{b}$, $m^{(2)}_{b}$ (Bob).

\item
They communicate these measurement outcomes 
to each other.
Then, they keep $\rho^{(0)}$
if  both of the measurement outcomes coincide
as $m^{(1)}_{a}=m^{(1)}_{b}$ and $m^{(2)}_{a}=m^{(2)}_{b}$. 
Otherwise, they discard $\rho^{(0)}$.  
\end{enumerate} 
Similarly to the single-selection protocol,
Alice and Bob iterate the procedures (ii)-(v)
by using the output states which survive the selection in (v)
as the input states for the next round
where the $ X $ and $ Z $ bases of their reference frames
are exchanged with a Hadamard transformation.

The action of these procedures 
with ideal local operations is described as
\begin{equation}
\begin{array}{c|cccc}
 \rho^{(0)} \diagdown \rho^{(1)} 
 & \phi^{(1)}_{0} & \phi^{(1)}_{1} & \phi ^{(1)}_{2} & \phi^{(1)}_{3}
 \\
 \hline
 \phi^{(0)}_{0} & \phi ^{(0)} _{0} & \times &  \times & \times
 \\
 \phi^{(0)}_{1} & \times & \phi^{(0)} _{1} &  \times  & \times
 \\
 \phi^{(0)}_{2} &  \times  & \phi ^{(0)} _{2} & \times & \times
 \\
 \phi^{(0)}_{3} &  \phi ^{(0)} _{3} & \times & \times & \times 
\end{array}
\nonumber
\end{equation}
with $ \rho^{(2)} = \phi^{(2)} _{0} $,
and so on with $ \rho^{(2)} = \phi^{(2)} _{1} ,
\phi^{(2)} _{2} , \phi^{(2)} _{3} $.
The double selection removes the errors more efficiently by using two ancilla EPR pairs.
For example, as seen in the above table for $\phi _{0}^{(0)} \otimes \phi _{3}^{(1)}$
with $\phi _{0}^{(2)}$,
we can detect and discard the $ Z $ error
propagation as $\phi _{0}^{(0)} \otimes \phi _{3}^{(1)} \rightarrow \phi _{3}^{(0)}$  
which is the first-order failure event in the single-selection scheme.
Thus, this double selection process significantly 
improves the maximum achievable fidelity of purified states.
The optimality of the double selection 
to detect the first-order errors will be discussed in detail later.

The purification map $\mathcal{D}$ of the double selection is descried as 
\begin{equation}
\mathbf{F}' =\mathcal{D}(\mathbf{F}),
\end{equation}
or in terms of  
a transition probability tensor $D_{i}^{jkl}$
as 
\begin{equation}
F'_{i} =\frac{1}{p_{\mathcal{D}}(\mathbf{F})}
\sum _{jkl} D_{i}^{jkl} F_{j}F_{k}F_{l},
\end{equation}
where 
\begin{equation}
p_{\mathcal{D}} (\mathbf{F})= \sum _{ijkl} D_{i}^{jkl} F_{j}F_{k}F_{l}
\label{success_prob_double}
\end{equation}
is the success probability responsible for the normalization.
The transition probability tensor $D_{i}^{jkl}$ is given in Appendix A.
\subsection{Performance analysis and comparison}
For each scheme,
we investigate the performance
by considering the working range of the local operations,
the maximum achievable fidelity of purified states,
the minimum fidelity required for the quantum communication channels
and the EPR resources consumed to achieve a target fidelity.
Here we assume that
the imperfect C-Not gate 
is implemented as a perfect C-Not gate
followed by two-qubit deporalizing errors with probabilities $p_{ij}$:
\begin{equation}
\mathcal{N} (\rho) 
= 
(1-p_{g}) \rho 
+ \sum _{(i,j)/(0,0)} 
p_{ij}
(\sigma _{i} \otimes \sigma _{j}) \rho (\sigma _{i} \otimes \sigma _{j}),
\end{equation}
where $i,j=0, 1, 2, 3$, and $p_{g}=\displaystyle\sum _{(i,j)/(0,0)} p_{ij}$.
The imperfect measurement in the $Z$ basis is described
with POVM elements and error probability $p_{m}$ as
\begin{eqnarray}
E_{0}&=& (1-p_{m})|0\rangle \langle 0| + p_{m} |1\rangle \langle 1|,
\\
E_{1} &=& (1-p_{m})|1\rangle \langle 1| + p_{m} |0\rangle \langle 0|.
\end{eqnarray}
The imperfect measurement in the $X$ basis is described  
similarly as
\begin{eqnarray}
 E_{+} &=& H E _{0} H, 
 \\
 E_{-} &=& H E_{1} H ,
 \end{eqnarray}
where $H$ denotes a Hadamard operation.
Then the transition probability tensors 
$S_{i}^{jk}$ and $D_{i}^{jkl}$,
which characterize the purification maps 
$\mathcal{S}$ and $\mathcal{D}$ respectively,
are calculated
including the error probabilities $p_{ij}$ and $p_{m}$ (see Appendix A).
By operating the purification map $\mathcal{A}$ recursively,
we can calculate the state-vector $\mathbf{F}^{(n)}$
at the $n$ round of purification iteration as 
\begin{eqnarray}
\mathbf{F}^{(n)} &=& \mathcal{A}(\mathbf{F}^{(n-1)})
\end{eqnarray}
($\mathcal{A=S}$ for the single selection and $\mathcal{D}$
for the double selection).
The EPR states are shared initially through 
the noisy communication channel $\mathcal{C}$
as
\begin{equation}
\mathcal{C}(\rho) = F_{\textrm{ch}} \rho 
+ \frac{1-F_{\textrm{ch}}}{3} 
\sum _{i=1}^{3} \sigma _{i} \otimes \sigma _{0}
\rho \sigma _{i} \otimes \sigma _{0}.
\end{equation}
Thus we set the initial state-vector for $\mathcal{C}( \phi _{0} )$ as
\begin{equation}
\mathbf{F}^{(0)} 
= \left( F_{\textrm{ch}}, \frac{1-F_{\textrm{ch}}}{3},
\frac{1-F_{\textrm{ch}}}{3},\frac{1-F_{\textrm{ch}}}{3}\right).
\end{equation}

\begin{figure}
\centering
\scalebox{1.3}{\includegraphics*[1.5cm,1.5cm][8cm,6.2cm]
{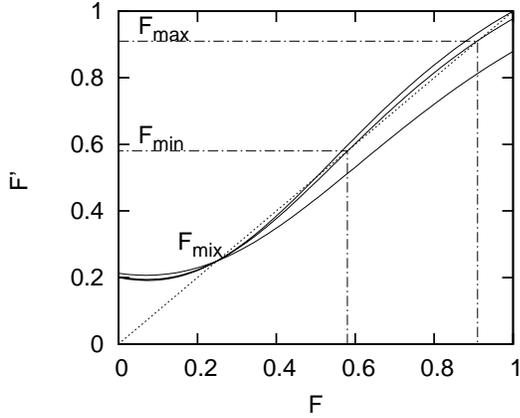}}
\caption{Purification curve. The fidelity of the output state $F'$
is plotted as a function of the fidelity of the input state $F$. 
Curves from top to bottom correspond to perfect local operations,
imperfect local operations inside the working range,
and those outside the working range.}
\label{PurificationCurve}
\end{figure}

In the case with ideal local operations,
the purification map $\mathcal{A}$ has 
three fixed points for $F \equiv F_{0}$ and $F' \equiv F'_{0} $,
i.e., the maximum achievable fidelity $F_{\textrm{max}}=1$,
the minimum required channel fidelity $F_{\textrm{min}}=1/2$ 
and the completely mixed state $F_{\textrm{mix}}=1/4$,
as seen from the purification curve in Fig. \ref{PurificationCurve}.
Then the imperfection of the local operations
shifts down the purification curve.
If the error probabilities $p_{g}$ and $p_{m}$ are inside the working range
of the purification scheme,
there still exist the 
three fixed points $F_{\textrm{max}}<1$, $F_{\textrm{min}}>1/2$
and $F_{\textrm{mix}}=1/4$. 
In this case, if the communication channel has a fidelity $F_{\textrm{ch}}$ 
higher than $F_{\textrm{min}}$,
we can achieve the fidelity $F_{\textrm{max}}$
by iterating the purification procedure.
On the other hand, outside the working range,
the purification procedure converts
the noisy EPR state to the
completely mixed state
($F_{\rm mix}=1/4$) with no entanglement. 
\begin{figure}
\centering
\scalebox{1.4}{\includegraphics*[1.5cm,1.5cm][8cm,6.2cm]{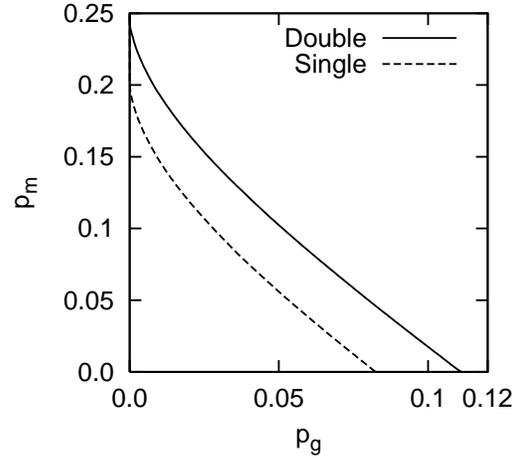}}
\caption{Working range $(p_{g},p_{m})$ of local operations. 
Each purification map $\mathcal{A}$ has 
the non-trivial fixed points $F_{\rm max}$ and $F_{\rm min}$ 
for the error probabilities $(p_{g},p_{m})$
below the threshold curve (solid line for $\mathcal{D}$
and dotted line for $\mathcal{S}$).}
\label{EPRRange}
\end{figure}
\begin{figure}
\scalebox{1.3}{\includegraphics*[1.7cm,1.5cm][8cm,6.2cm]{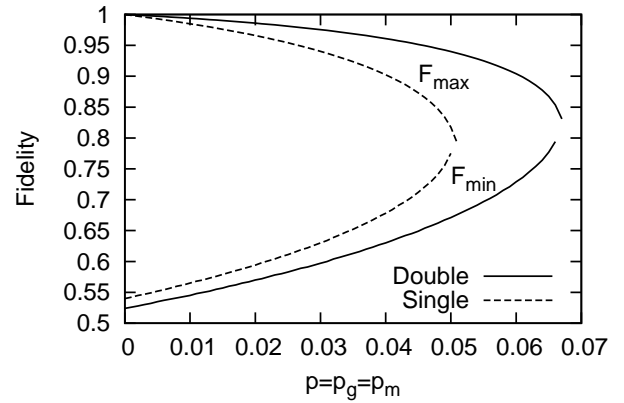}}
\caption{Maximum achievable fidelity $F_{\textrm{max}} $ (upper curves) 
and minimum required channel fidelity $ F_{\textrm{min}} $ (lower curves)
are plotted as functions of the error probability $ p=p_{g}=p_{m} $
for the double and single selections.}
\label{EPRFidelity}
\end{figure}

The working range $(p_{g},p_{m})$
of the local operations
is shown in Fig. \ref{EPRRange}
for each purification scheme.
Here the error probabilities of the C-Not gate
are taken equally as $p_{ij}=p_g /15$.
As seen in Fig. \ref{EPRRange},
the double-selection scheme
has higher thresholds for the error probabilities
than the single-selection scheme.
Thus it works well with noisier local operations.
If we choose a different error distribution
$p_{i0}=p_{0i}=q_{i}$, $p_{ij}=q_{i}q_{j}$ ($i \neq j$ and $i, j \neq 0$)
and $p_{g}=\displaystyle\sum _{i=1}^{3} q_{i}$, 
as adopted in Ref. \cite{Kay08},
we estimate 
the threshold values $3-4 \%$ and $4-5\%$ of $p_{g}$ with $p_{m}=0$ 
for the single and double selections, respectively,
which are consistent with an upper bound $5.3 \%$
derived in Ref. \cite{Kay08}.

In Fig. \ref{EPRFidelity}, $ F_{\textrm{max}} $ (upper curves) 
and $F_{\textrm{min}}$ (lower curves)
are plotted as functions of the error probability $p$
($p=p_{g}=p_{m}$ for definiteness).
The double selection achieves a higher fidelity of purification
$ F_{\textrm{max}} $
with a less fidelity of communication channel $ F_{\textrm{min}} $,
compared with the single selection.
These improvements are due to the fact that 
the double selection can detect more efficiently the first-order errors,
as discussed in detail later.

\begin{figure}
\scalebox{1.2}{\includegraphics*[1.7cm,1.5cm][10cm,6.5cm]{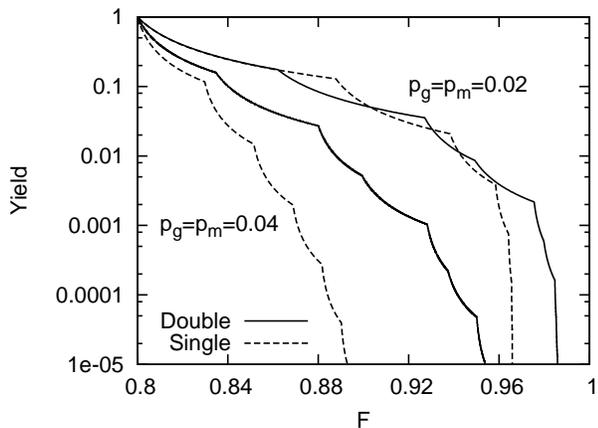}}
\caption{The yield $Y_{\mathcal{A}}(F,F_{\rm ch}=0.8)$ is plotted 
as a function of the target fidelity $F$ for each protocol 
with $p_{g}=p_{m}=0.02$ (upper curves) 
and $p_{g}=p_{m}=0.04$ (lower curves).}
\label{EPRYield}
\end{figure}
We compare the yields  
of the purification protocols.
The yield $Y_{\mathcal{A}}(F,F_{\rm ch})$ is defined as
the inverse of the number of EPR pairs
consumed to achieve a target fidelity $F$
under the channel fidelity $F_{\textrm{ch}}$
\cite{Kruszynska06b}. 
It is calculated for each scheme $\mathcal{A}=\mathcal{S},\mathcal{D}$ as
\begin{equation}
Y_{\mathcal{A}} (F,F_{\textrm{ch}}) 
=
\left[ \prod _{n=1}^{n_{\mathcal{A}}(F,F_{\textrm{ch}})} 
N_{\mathcal{A}}/p_{\mathcal{A}}(\mathbf{F}^{(n-1)}) \right] ^{-1},
\end{equation}
where $n_{\mathcal{A}}(F,F_{\textrm{ch}})$
denotes the minimum number of rounds 
which is required to achieve the fidelity $F$,
$p_{\mathcal{A}}(\mathbf{F}^{(n-1)})$ denotes 
the probability to pass the purification procedure 
at the round $n$, as given in Eqs. (\ref{success_prob_single})
and (\ref{success_prob_double}),
and $N_{\mathcal{A}}$ denotes the number of EPR pairs
consumed at each round ($N_{\mathcal{S}}=2$ and $N_{\mathcal{D}}=3$).
We plot in Fig. \ref{EPRYield} the yield
$Y_{\mathcal{A}}(F,F_{\rm ch}=0.8)$ 
as a function of the target fidelity $F$
for each protocol with $p_g=p_m = 0.02$ (upper curves) and 
$p_g=p_m = 0.04$ (lower curves).
With less noisy local operations ($p_g=p_m=0.02$),
both schemes provide comparable yields
to achieve a target fidelity $F \approx 0.9$, 
where the numbers of rounds are $n_{\mathcal{S}}=4$ (single) 
and $n_{\mathcal{D}}=2$ (double).
On the other hand, 
even when noisier local operations ($p_g=p_m=0.04$) are used,
the double-selection scheme still provides a reasonable yield
to achieve a target fidelity $F \approx 0.9$,
where $n_{\mathcal{S}}=16$ and $n_{\mathcal{D}}=4$.
Since the double-selection scheme uses 
three EPR pairs at each round,
it may be thought to cost more resources
than the single-selection scheme
with two EPR pairs at each round.
However, this is not the case.
As shown in the above,
the double-selection scheme provides a comparable or even better yield.
This is because
the double-selection scheme
increases the fidelity of EPR pairs
considerably faster than the single-selection scheme.
In the case of the double selection,
an additional EPR pair is worth enough
to enhance the fidelity of purified states,
as discussed in the following.

\subsection{Optimality of the double selection}
\begin{figure}
\centering
\scalebox{.43}{\includegraphics*[0cm,0cm][20cm,11cm]{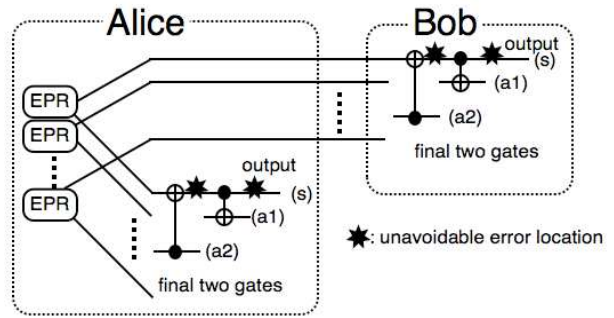}}
\caption{Setup of a purification protocol.
An upper bound on the fidelity is determined in the first order
by the unavoidable errors (indicated by black stars)
introduced by the final two C-Not gates.
Similar bounds are obtained with other configurations
of two-qubit gates.}
\label{upperbound}
\end{figure}

We here discuss the optimality of the double selection
in the first order of the errors.
We first consider an upper bound on the fidelity of purified states
by general arguments.
Then, it will be shown that the double selection
really saturates this upper bound in the low noise regime.

While purification protocols contain many imperfect local operations,
we can reasonably derive an upper bound on the fidelity
in the first order by noting the errors
introduced by the final two gate operations
applied on the source EPR pair.
Here we assume that the local operations for purification
are implemented by bilateral two-qubit Clifford gates
and measurements.
(One-qubit gates may be absorbed into relevant two-qubit gates
and local reference frames.)
Specifically, consider a purification protocol
with a combination of two bilateral C-Not gates at the final step,
as shown in Fig. \ref{upperbound}.
We inspect these final C-Not gate operations in the Alice's site
to observe the undetectable errors
left on the output state of the source qubit (s).
(The same argument is made in the Bob's site.)
Passing through this set of the final gates,
the preceding errors on the source qubit (s) are propagated
to either or both of the ancilla qubits (a1) and (a2).
Thus these preceding errors are all regarded to be detectable, that is
an optimal purification protocol is expected to remove them
by post-selection after the measurements of the ancillas.
The upper bound on the fidelity is rather determined
by some of the errors introduced by the final two gates themselves
(black stars in Fig. \ref{upperbound})
which are undetectable and thus unavoidable
without leaving any information on the ancillas.
As for the second final C-Not gate,
only the $ \sigma^{\rm (s)}_{3} \otimes \sigma^{\rm (a2)}_{0} $ error
with the probability $ p_{30} $ is undetectable,
since it commutes with the final C-Not gate.
On the other hand,
all the $ \sigma^{(\rm s)}_{i} \otimes \sigma^{\rm (a1)}_{0} $ errors
of the final C-Not gate with the probabilities $ p_{i0} $
are undetectable inevitably, since the output source qubit
does not interact with ancillas afterward.
By adding these undetectable errors due to the final two C-Not gates
in the sites of Alice and Bob,
the upper bound on the fidelity is placed
in the first order of the errors as
\begin{eqnarray}
F_{\rm upper} = 1 - 2 \left( p_{30} + \sum_{i=1}^3 p_{i0} \right)
- \mathcal{O} ( p_g^2 ) .
\label{upperboundA}
\end{eqnarray}
Note here that the measurement error is not involved
in considering the detectability of the errors
with their propagation to the ancilla qubits.
A portion of the right output state $ \phi_{0} $
may be discarded by the errors in measuring the ancillas
for verification.  This slight reduction of the right state
is cancelled by the renormalization after the post-selection
except for the higher-order contributions.
Thus the measurement error does not contribute
to the upper bound in the first order.
The precise upper bound will be lower than this first-order bound
due to the higher-order contributions of the gate and measurement errors,
though it is beyond our scope to estimate them analytically.

Another protocol may be considered by exchanging the final two C-Not gates
in Fig. \ref{upperbound}.
(This is actually the case in the recurrence protocol considered so far
when the purification procedure is finished at an even round.)
Similarly, by observing the undetectable errors
introduced by the final two C-Not gates,
the upper bound on the fidelity is placed as
\begin{eqnarray}
F'_{\rm upper} = 1 - 2 \left( p_{10} + \sum_{i=1}^3 p_{i0} \right)
- \mathcal{O} ( p_g^2 ) .
\label{upperboundB}
\end{eqnarray}
We note that these upper bounds
in Eqs. (\ref{upperboundA}) and (\ref{upperboundB}) are valid
even if the final two C-Not gates are set in the same direction
(e.g., both the C-Not gates are controlled by the source qubit).
In such configurations, some of the preceding errors on the source qubit
become undetectable, commuting with the final two gates.
Thus these configurations merely lower the fidelity.

These two upper bounds $ F_{\rm upper} $ and $ F'_{\rm upper} $
coincide with each other for the uniform distribution of the gate errors
$ p_{ij} = p_g / 15 $,
as adopted in calculating the maximum achievable fidelities
of the single and double selections (Fig. \ref{EPRFidelity}).
In general, the upper bound is given
by $ {\rm max} [ F_{\rm upper} , F'_{\rm upper} ] $
depending on the error distribution;
in a recurrence protocol we should determine
whether the purification procedure is finished at an even or odd round.
We may also use other two-qubit Clifford gates instead of C-Not gates.
Then, we obtain a similar upper bound
with a suitable permutation among $ p_{i0} $'s
in Eq. (\ref{upperboundA}) or Eq. (\ref{upperboundB}),
by considering which error of the second final gate
is left commuting with the final gate.

We now show that the double selection really removes
all the detectable errors in the first order,
achieving the upper bound $ F_{\rm upper} $.
(We here adopt the uniform error distribution for definiteness.)
The recurrence protocol considered so far,
with either single or double selection,
has the setup as shown in Fig. \ref{upperbound}
with the proper exchange of the reference frames at each round.
In the single selection (Fig. \ref{single}),
the $ \sigma_{1} $ and $ \sigma_{2} $
($ \sigma_{2} $ and $ \sigma_{3} $) errors
on each ancilla EPR pair are detected by the $ Z $ ($ X $) measurement,
while the $ \sigma_{3} $ ($ \sigma_{1} $) error is not detected.
In the double selection (Fig. \ref{double}),
the primary ancilla EPR pair is further verified
with the secondary ancilla EPR pair
so that all the errors on the primary ancilla EPR pair
are detected in the first order.
That is, all the detectable errors on the source EPR pair
are certainly removed by the double selection
through the interaction with the primary ancilla EPR pairs,
as expected in deriving the upper bound.
Therefore, the double-selection protocol achieves the upper bound
in the first order as
\begin{eqnarray}
F_{\rm max}^{\mathcal D}
&=& F_{\rm upper} - \mathcal{O}(p_{g}^{2}) .
\end{eqnarray}
This has been really confirmed by the numerical calculation
in the low gate noise regime $ p_{g} < 0.02 $,
almost independently of the measurement error $ p_{m} < 0.05 $.

On the other hand, as noted above,
in the single selection (Fig. \ref{single})
the $ \sigma_{3} $ ($ \sigma_{1} $) error on the ancilla EPR pair
cannot be detected by the $ Z $ ($ X $) measurement at an odd (even) round.
Even in this case, all the errors before the second final C-Not gate
are detected in the first order through the final two rounds,
because the reference frames are properly exchanged at each round.
However, some of the errors which have been excluded as detectable
in deriving $ F_{\rm upper} $ are actually left
in the single-selection protocol.
By counting these extra errors due to the final two C-Not gates
in Fig. \ref{upperbound}, we find that the maximum achievable fidelity
of the single selection decreases as
\begin{eqnarray}
F_{\rm max}^{\mathcal S} =  F_{\rm max}^{\mathcal D} - (8/15) p_{g}
- \mathcal{O}(p_{g}^{2}) .
\end{eqnarray}

As seen so far, the double selection is necessary and sufficient
to remove fully the detectable first-order errors,
saturating the upper bound on the fidelity in the low noise regime.
There is no room for the triple (or more) selection to improve the fidelity
except for the higher-order error contributions.
In practice, as the triple selection itself
introduces more errors with additional noisy operations,
it will hardly improve the maximum achievable fidelity
of the double selection.

\section{Multi-partite entanglement purification} 
\label{Multi-partite}
Recently entanglement purification is applied
to a large class of multi-partite entanglements
including two-colorable graph states
\cite{ADB0305,Kruszynska06a,Miyake05}.
We can extend the double-selection scheme
to such protocols for multi-partite entanglement purification.
Specifically, we here consider the purification
of two-colorable graph states \cite{ADB0305}.

A graph state is expressed as an eigenstate of
a set of stabilizer operators
\begin{equation}
K_{j} = X_{j} \bigotimes_{k \in V_{j}} Z_{k} ,
\end{equation}
where $ X_{j} $ and $ Z_{k} $ denote the Pauli operators
acting on the $ j $-th and $k$-th  qubits respectively,
and $ V_{j} $ denotes the set of vertices
which are connected to the $ j $-th qubit \cite{Dur07}.
Then a graph state is described as
\begin{equation}
K_{j} | \Psi_{ \mu _{1} \mu _{2} \cdots \mu_{N}} \rangle
= (-1)^{\mu_{j}} | \Psi_{ \mu_{1} \mu_{2} \cdots \mu_{N}} \rangle ,
\end{equation}
where $ (-1)^{\mu _{j}} (\mu _{j} = 0,1) $ 
denotes the eigenvalue $\pm 1$
of the operator $ K_{j} $.
Especially a graph state which can be divided into two partitions $A$ and $B$
such that no vertices within one set are connected by edges
is called a two-colorable graph state.  
It is described as
\begin{equation}
| \Psi_{\mbox{\boldmath $\mu$}_{A}, \mbox{\boldmath $\mu$}_{B}} \rangle ,
\end{equation} 
where $ \mbox{\boldmath $\mu$}_{A} $ and $ \mbox{\boldmath $\mu$}_{B} $
denote the eigenvalues of the vertices
within the partitions $ A $ and $ B $, respectively.

\begin{figure}
\centering
\scalebox{.45}{\includegraphics*[0cm,0cm][20cm,8cm]{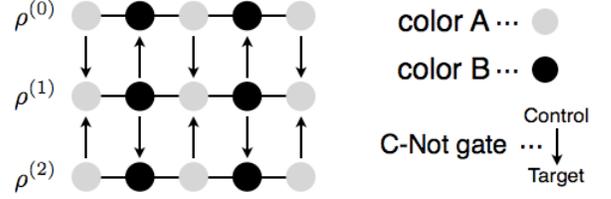}}
\caption{Purification of two-colorable graph states with double selection.}
\label{protocol2}
\end{figure}

The entanglement purification protocol with double selection
for a noisy mixture of $n$-partite two-colorable graph states
\begin{equation}
\rho =  \sum _{\mbox{\boldmath $\mu$}_{A}, \mbox{\boldmath $\mu$}_{B}}
\lambda _{\mbox{\boldmath $\mu$}_{A}, \mbox{\boldmath $\mu$}_{B}}
|\Psi _{\mbox{\boldmath $\mu$}_{A}, \mbox{\boldmath $\mu$}_{B}} \rangle
\langle \Psi _{\mbox{\boldmath $\mu$}_{A}, \mbox{\boldmath $\mu$}_{B}}|
\end{equation}
is implemented as follows
(see Fig. \ref{protocol2}):
\begin{enumerate}[(i)]
\item
Alice, Bob, $\cdots$, Nancy
share three identical two-colorable graph states
$ \rho^{(0)} $, $ \rho^{(1)} $ and $ \rho^{(2)} $
through a noisy quantum channel.
\item
They operate a multi-lateral C-Not gate
for $ \rho^{(0)} $ and $ \rho^{(1)} $
with the partition $A$ [$B$] of $ \rho^{(0)} $ [$ \rho^{(1)} $] as the control
and the partition $A$ [$B$] of $ \rho^{(1)} $ [$ \rho^{(0)} $] as the target,
respectively.

\item
Similarly they operate a multi-lateral C-Not gate
for $ \rho^{(2)} $ and $ \rho^{(1)} $.

\item
They multi-laterally measure the partition $A$ [$B$] of $ \rho^{(1)} $
and the partition $A$ [$B$] of $ \rho^{(2)} $
in the $ Z $ and $ X $ [$ X $ and $ Z $] bases respectively,
and obtain measurement outcomes $(-1)^{\xi^{(1)} _{i}}$
[$(-1)^{\zeta ^{(1)} _{j}}$]
and $(-1)^{\xi^{(2)} _{i}}$ [$(-1)^{\zeta ^{(2)} _{j}}$],
where $\xi _{i}, \zeta _{j} = 0,1$, 
and the qubit $i$ [$j$] belongs to the partition $A$ [$B$].

\item
They communicate these measurement outcomes to each other.
Then they keep $ \rho^{(0)}$ 
if for all $i$ and $j$,
$ \zeta^{(1)}_{j} \oplus \sum_{k \in V_{j}} \xi^{(1)}_{k}=0$
and
$ \xi^{(2)}_{i} \oplus \sum_{k \in V_{i}} \zeta^{(2)}_{k}=0$,
which implies
$ \mbox{\boldmath{$\mu$}}^{(0)} _ A 
\oplus \mbox{\boldmath{$\mu$}}^{(1)} _ A \oplus 
 \mbox{\boldmath{$\mu$}}^{(2)} _ A= \mathbf{0}$
and
$\mbox{\boldmath{$\mu$}}^{(1)} _ B 
\oplus \mbox{\boldmath{$\mu$}}^{(2)} _ B= \mathbf{0}$,
respectively,
where $\oplus$ denotes bitwise addition modulo 2.
\end{enumerate}
They iterate the procedures (ii)-(v)
by using the output states which survive the selection in (v)
as the input  states for the next round where
the $ X $ and $ Z $ bases of their reference frames
are exchanged with a Hadamard transformation,
similarly to the bipartite case.
These procedures with perfect operations act as
\begin{eqnarray}
| \Psi^{(0)}_{\mbox{\boldmath $\mu$}^{(0)}_{A},
\mbox{\boldmath $\mu$}^{(0)}_{B}} \rangle
&\rightarrow&
| \Psi^{(0)}_{\mbox{\boldmath $\mu$}^{(0)}_{A},
\mbox{\boldmath $\mu$}^{(0)}_{B}
\oplus \mbox{\boldmath $\mu$}^{(1)}_{B}} \rangle,
\\
| \Psi^{(1)}_{\mbox{\boldmath $\mu$}^{(1)}_{A},
\mbox{\boldmath $\mu$}^{(1)}_{B}} \rangle
&\rightarrow&
| \Psi^{(1)}_{\mbox{\boldmath $\mu$}^{(0)}_{A}
\oplus \mbox{\boldmath $\mu$}^{(1)}_{A}
\oplus \mbox{\boldmath $\mu$}^{(2)}_{A},
\mbox{\boldmath $\mu$}^{(1)}_{B}} \rangle,
\\
| \Psi^{(2)}_{\mbox{\boldmath $\mu$}^{(2)}_{A},
\mbox{\boldmath $\mu$}^{(2)}_{B}}\rangle
&\rightarrow&
| \Psi^{(2)}_{\mbox{\boldmath $\mu$}^{(2)}_{A},
\mbox{\boldmath $\mu$}^{(1)}_{B}
\oplus \mbox{\boldmath $\mu$}^{(2)}_{B}}\rangle.
\end{eqnarray}
Thus in the double-selection scheme we can detect even the propagation of the errors
$ \mbox{\boldmath $\mu$}^{(1)}_{B} $
from $ \rho^{(1)} $ to $ \rho ^{(0)}$ by using $ \rho^{(2)} $ the same as the bipartite case.

As an example 
of the multi-partite purification 
with the double selection,
we numerically investigate a CSS code state, 
specifically the Steane's 7-qubit code state,
which is a two-colorable graph state.
Then its performance is
compared with the ADB protocol \cite{ADB0305}.
We consider 
the multiparty communication situation, where $n$-qubit two-colorable graph states
$| \Psi _ {\mathbf{0}_{A},\mathbf{0}_{B}} 
\rangle \langle \Psi _{\mathbf{0}_{A},\mathbf{0}_{B}}|$
are shared through $n$ identical noisy channels $\mathcal{C}^{\otimes n}$.
The noisy copies $\rho _{\rm in}= \mathcal{C}^{\otimes n}
(| \Psi _{\mathbf{0}_{A},\mathbf{0}_{B}} 
\rangle \langle \Psi _{\mathbf{0}_{A},\mathbf{0}_{B}}|)$
of the $n$-qubit entangled state
are purified with
the noisy C-Not gates and measurements
the same as the bipartite case.
We simulate directly the noisy operations
in the communication and purification procedures 
by using Monte-Carlo method.
(It is highly complicated to provide the purification map 
in terms of the transition probability tensor.)
The fidelity 
of the purified state $\rho '$ is measured by
\begin{equation}
F(\rho',|\Psi _{\mathbf{0}_{A},\mathbf{0}_{B}} \rangle)
=
\langle \Psi _{\mathbf{0}_{A},\mathbf{0}_{B}} |
\rho ' |\Psi _{\mathbf{0}_{A},\mathbf{0}_{B}} \rangle. 
\end{equation}
\begin{figure}
\scalebox{1.2}{\includegraphics*[1.7cm,1.5cm][9cm,6.5cm]
{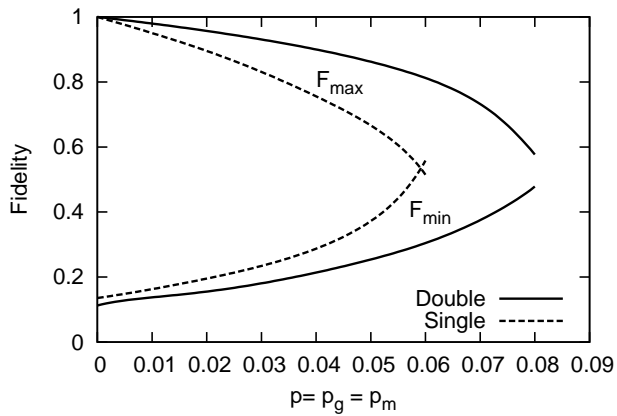}}
\caption{Maximum achievable fidelity $F_{\textrm{max}}$ (upper curves)
and minimum required channel fidelity $F_{\textrm{min}}$ (lower curves)
for the Steane's 7-qubit code state $ | 0_{L} \rangle $
are plotted as functions of the error probability $ p=p_{g}=p_{m} $
for the single and double selections.}
\label{MultiCommu}
\end{figure}
\begin{figure}
\scalebox{1.2}{\includegraphics*[1.7cm,1.5cm][10cm,7cm]{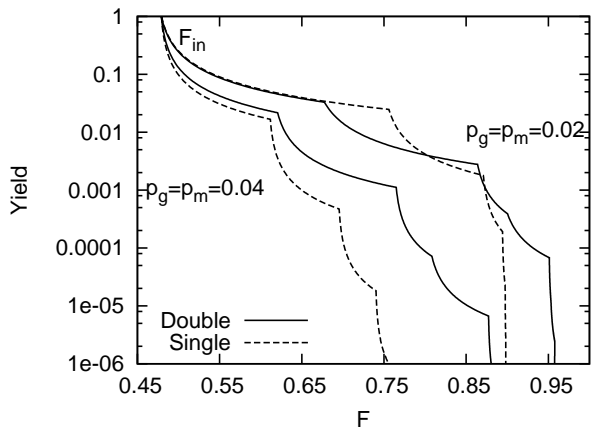}}
\caption{The yield $Y_{\mathcal{A}}(F,F_{\rm ch}=0.9)$ is plotted 
as a function of the target fidelity $F$
for each protocol with
$p_{g}=p_{m}=0.02$ (upper curves) 
and $p_{g}=p_{m}=0.04$ (lower curves).}
\label{CSSYield}
\end{figure}
If the initial fidelity 
\begin{eqnarray}
F_{\rm in} \equiv F(\rho _{\rm in}, |\Psi _{\mathbf{0}_{A},\mathbf{0}_{B}} \rangle)
= F_{\rm ch}^{7} + \mathcal{O}((1-F_{\rm ch})^{3})
\end{eqnarray}
is higher than $F_{\rm min}$,
we can achieve the fidelity $F_{\rm max}$ by iterating the purification procedure.

The resultant maximum achievable fidelity $F_{\textrm{max}}$
and minimum required initial fidelity $F_{\textrm{min}}$
are plotted for the Steane's 7-qubit code state 
$|\Psi _{\mathbf{0}_{A},\mathbf{0}_{B}} \rangle=|0_{L}\rangle$
as functions of the error probability $p=p_{g}=p_{m}$
in Fig. \ref{MultiCommu}.
As we expect,
the double-selection protocol 
achieves considerably 
higher fidelity of the purified states
requiring lower fidelity of the channels.
Then the noise threshold for the local operations
is improved from $5-6\%$ (single) to $8-9\%$ (double).
It is also seen in Fig. \ref{CSSYield} that
both schemes provide comparable yields,
similarly to the bipartite case.
The yields are, however, significantly low compared with the bipartite case.
This is because 
the coincidence of the more measurement outcomes is required 
in the multi-partite case,
reducing the success probability of post-selection.
The upper bound on the fidelity with imperfect local operations
can be extend to the two-colorable graph states in the same line.
Then the maximum achievable fidelity $F_{\rm max}$ of the double-selection scheme
saturates the upper bound $F_{\rm upper}=1-7 \times (4/15) p_{g}$ for $n=7$ with $p_{ij}=p_{g}/15$,
which is really confirmed in the low noise regime by the numerical simulation.

The two-colorable graph states,
including cluster and CSS code states,
play very important roles
in quantum computation as well as quantum communication.
Then these results really indicate that 
the double selection is profitable
also in quantum computation.
In fact,
encoded ancilla qubits are used to stabilize a computation
in a fault-tolerant way,
and the performance of computation highly depends on the fidelity 
of these ancilla qubits \cite{Knill05,Eastin07}.
In the usual fault-tolerant context \cite{Steane97,Knill05,Cross07},
these encoded ancilla qubits 
are prepared through the single selection.
Thus the double selection has a good potential to improve
the noise threshold of fault-tolerant computation.
The verification process with the double selection
is used in fault-tolerant computation
with concatenated construction of verified logical cluster states,
achieving a considerably high noise threshold $ \sim 4\% $ \cite{FY06}.
We will discuss elsewhere
further applications of the double selection 
for quantum computation.

\section{Conclusion}
\label{conclusion}
We have investigated
entanglement purification with double selection.
It has been shown that the maximum achievable fidelity 
can be improved significantly by the double selection
under noisy communication channels and imperfect local operations.
Furthermore 
the double-selection scheme provides
a  reasonable yield, which is comparable or even better than the single-selection scheme.
The double selection is really optimal
to remove the first-order errors,
achieving the upper bound on the fidelity 
of purified states in the lower noise regime. 
These results really indicate that
the double selection is  
more suitable for realization 
of entanglement-based protocols.
The double selection is also applied
to multi-partite entanglement purification,
specifically two-colorable graph states.
The improvement of the fidelity has been shown numerically
for the Steane's 7-qubit code state.
The double selection can be extended to all graph states in the same way.
Since multi-partite entangled states,
such as CSS codes and cluster states,
play very important roles in quantum computation
as well as quantum communication,
the double selection  
has a good potential to improve the performance of quantum computation.

\begin{acknowledgments}
This work was supported by 
JSPS Research Fellowships for Young Scientists No. 20$\cdot$2157.
\end{acknowledgments}

\appendix
\section{Derivation of Transition Probability Tensors}
The purification procedure of the single selection
before the post-selection is described
as a linear map $ {\tilde {\mathcal S}} $
of the two Bell states $ \phi^{(0)}_{i} $ and $ \phi^{(1)}_{j} $ as
\begin{equation}
{\tilde {\mathcal S}} ( \phi^{(0)}_{i} \otimes \phi^{(1)}_{j} )
= \tilde{S}^{ij}_{kl} \phi^{(0)}_{k} \otimes \phi^{(1)}_{l} .
\end{equation}
This map consists of the noisy bilateral C-Not gate
$ {\mathcal G} ( \phi^{(0)}_{i} \otimes \phi^{(1)}_{j} )
= G^{ij}_{ab} \phi^{(0)}_{a} \otimes \phi^{(1)}_{b} $,
the bilateral $ Z $-basis measurement with error
$ {\mathcal M} ( \phi^{(1)}_{b} ) = M^{b}_{l} \phi^{(1)}_{l} $
and the exchange of the reference frames at each round
by the bilateral $ H $ transformation (ideal for simplicity)
$ {\mathcal H} ( \phi^{(0)}_{a} ) = H^{a}_{k} \phi^{(0)}_{k} $:
\begin{equation}
\tilde{S}^{ij}_{kl} = H^{a}_{k} {M}^{b}_{l} G^{ij}_{ab} .
\end{equation}
The noisy bilateral C-Not gate $ G^{ij}_{km} $ is decomposed
into the ideal bilateral C-Not gate $ U^{ij}_{ab} $
and the bilateral combination $ N^{cd}_{km} N^{ab}_{cd} $
of the C-Not gate noises at Alice and Bob:
\begin{equation}
G^{ij}_{km} = N^{cd}_{km} N^{ab}_{cd} U^{ij}_{ab} .
\end{equation}
Note here that the Bell states are two-colorable graph states
up to a local Hadamard operation as
\begin{equation}
| \phi_{i} \rangle
= (I \otimes H) | \Psi_{\mu^{i}_{A}, \mu^{i}_{B}} \rangle ,
\end{equation}
where
\begin{equation}
\mbox{\boldmath $\mu$}^{i}
= ( \mu^{i}_{A}, \mu^{i}_{B} ) = (0,0), (1,0), (1,1), (0,1)
\end{equation}
for $ i = 0, 1, 2, 3$, respectively.
In the graph state representation,
the action of the ideal bilateral C-Not gate is described as
\begin{equation}
| \Psi^{(0)}_{\mu^{i}_{A}, \mu^{i}_{B}} \rangle
| \Psi^{(1)}_{\mu^{j}_{A}, \mu^{j}_{B}} \rangle
\rightarrow
| \Psi^{(0)}_{\mu^{i}_{A}, \mu^{i}_{B} \oplus \mu^{j}_{B}} \rangle
| \Psi^{(1)}_{\mu^{i}_{A} \oplus \mu^{j}_{A}, \mu^{j}_{B}} \rangle ,
\end{equation}
or simply
\begin{equation}
{\tilde{\mathcal U}} ( \mbox{\boldmath $\mu$}^{i}
\otimes \mbox{\boldmath $\mu$}^{j} )
= ( \mu^{i}_{A}, \mu^{i}_{B} \oplus \mu^{j}_{B} )
\otimes ( \mu^{i}_{A} \oplus \mu^{j}_{A}, \mu^{j}_{B} ) .
\end{equation}
Then the transition tensor of the ideal bilateral C-Not gate
is given by
\begin{equation}
U^{ij}_{ab} = \left\{ \begin{array}{cl}
1 & [ \mbox{\boldmath $\mu$}^{i}
\otimes \mbox{\boldmath $\mu$}^{j}
= {\tilde{\mathcal U}} ( \mbox{\boldmath $\mu$}^{a}
\otimes \mbox{\boldmath $\mu$}^{b} ) ] \\
0 & [ \mbox{\boldmath $\mu$}^{i}
\otimes \mbox{\boldmath $\mu$}^{j}
\not= {\tilde{\mathcal U}} ( \mbox{\boldmath $\mu$}^{a}
\otimes \mbox{\boldmath $\mu$}^{b} ) ]
\end{array} \right. ,
\end{equation}
e.g., $ U^{13}_{22} = 1 $
for $ \mbox{\boldmath $\mu$}^{2} \otimes \mbox{\boldmath $\mu$}^{2}
= (1,1) \otimes (1,1)
\rightarrow (1,0) \otimes (0,1)
= \mbox{\boldmath $\mu$}^{1} \otimes \mbox{\boldmath $\mu$}^{3} $
by $ {\tilde{\mathcal U}} $, and so on.
The noise on the C-Not gate is given by
\begin{equation}
N^{ab}_{cd} = p_{ij}
[ (ac) \in {\mathcal P}_{\sigma_{i}}, (bd) \in {\mathcal P}_{\sigma_{j}} ] ,
\end{equation}
e.g., $ N^{00}_{10} = p_{10} $ for $ (ac) = (01) $
with $ \sigma_{1} \otimes \sigma _{0} \phi_{0} \sigma_{1} \otimes \sigma _{0}= \phi_{1} $
and $ (bd) = (00) 
$ with $ \sigma_{0} \otimes \sigma _{0} \phi_{0} \sigma_{0} \otimes \sigma _{0} = \phi_{0} $,
and so on,
where $ \displaystyle{p_{00}
= 1 - \sum _{(i,j)/(0,0)} p_{ij} = 1 - p_{g}} $,
and $ \sigma_{i} $'s provide the permutations
among the Bell states $ \phi_{i} $'s as
\begin{eqnarray}
{\mathcal P}_{\sigma_{0}} &=&
\left( \begin{array}{cccc} 0 & 1 & 2 & 3
\\ 0 & 1 & 2 & 3 \end{array} \right) ,
\\
{\mathcal P}_{\sigma_{1}} &=&
\left( \begin{array}{cccc} 0 & 1 & 2 & 3
\\ 1 & 0 & 3 & 2 \end{array} \right) ,
\\
{\mathcal P}_{\sigma_{2}} &=&
\left( \begin{array}{cccc} 0 & 1 & 2 & 3
\\ 2 & 3 & 0 & 1 \end{array} \right) ,
\\
{\mathcal P}_{\sigma_{3}} &=&
\left( \begin{array}{cccc} 0 & 1 & 2 & 3
\\ 3 & 2 & 1 & 0 \end{array} \right) .
\end{eqnarray}
The bilateral $ Z $-basis measurement is given by
\begin{equation}
M^{b}_{l} = m^{e}_{l} m^{b}_{e}
\end{equation}
as a product of the single measurements given by
\begin{equation}
m^{b}_{e} = \left\{ \begin{array}{ll}
1 - p_{m} & [ (be) \in {\mathcal P}_{\sigma_{0}} ] \\
p_{m} & [ (be) \in {\mathcal P}_{\sigma_{1}} ] \\
0 &  [ (be) \in {\mathcal P}_{\sigma_{2}} , {\mathcal P}_{\sigma_{3}} ]
\end{array} \right. .
\end{equation}
The bilateral Hadamard operation is given by
\begin{equation}
H^{a}_{k} = h^{e}_{k} h^{a}_{e} ,
\end{equation}
where $ h^{0}_{0} = h^{1}_{3} = h^{2}_{2} = h^{3}_{1} = 1 $
and $ h^{a}_{e} = 0 $ for the others.
After all the transition probability tensor $ S^{jk}_{i} $
of the single selection is obtained
by picking up the post-selected states with $ l = 0 , 3 $
from $ \tilde{S}^{jk}_{il} $ as
\begin{equation}
S^{jk}_{i} = \tilde{S}^{jk}_{i0} + \tilde{S}^{jk}_{i3} .
\end{equation}

Similarly, the purification procedure of the double selection
before the post-selection is described as a linear map
$ {\tilde {\mathcal D}} $ of the three Bell states
$ \phi^{(0)}_{i} $, $ \phi^{(1)}_{j} $, $ \phi^{(2)}_{k} $ as
\begin{equation}
\tilde{D}^{ijk}_{lmn}
= H^{a}_{l} M^{c}_{m} \tilde{M}^{d}_{n} G^{kb}_{dc} G^{ij}_{ab} ,
\end{equation}
where
\begin{equation}
\tilde{M}^{d}_{n} = H^{f}_{n} M ^{e}_{f} H^{d}_{e}
\end{equation}
provides the $ X $-basis measurement.
Then the transition probability tensor of the double selection
is obtained by the post-selection as
\begin{equation}
D^{jkl}_{i} =  \sum_{m=0,3 ; n=0,1} \tilde{D}^{jkl}_{imn} .
\end{equation}

\end{document}